\UseRawInputEncoding
\documentclass{article}
\pdfoutput=1 
\usepackage{caption}
\usepackage{textcomp}
\usepackage{amsfonts}
\usepackage{amsmath}
\usepackage{amsthm}
\usepackage{enumitem}
\usepackage{amssymb,bm}
\usepackage[mathscr]{eucal}
\usepackage{graphicx}
\usepackage{color}
\usepackage{cite}
\usepackage{comment}
\usepackage{geometry}
\usepackage{multirow}
\usepackage{subcaption}
\usepackage{float}
\usepackage{booktabs}
\usepackage{algorithm} 
\usepackage{hyperref}
\usepackage{algorithmic} 
\usepackage{makecell} 
\usepackage{pifont} 

\usepackage{CJK}
\usepackage[utf8]{inputenc}
\usepackage{authblk} 

\title{\bf HDKD: Hybrid Data-Efficient Knowledge Distillation Network for Medical Image Classification}
\begin{document}
\author{Omar S. EL-Assiouti\textsuperscript{*}}
\author{Ghada Hamed}
\author{Dina Khattab}
\author{Hala M. Ebied}

\affil{\small Department of Scientific Computing, Ain Shams University, Cairo 11566, Egypt.}

\renewcommand{\thefootnote}{\fnsymbol{footnote}}
\footnotetext[1]{Corresponding author at: Department of Scientific Computing, Ain Shams University, Cairo 11566, Egypt. \\
E-mail address: omarsherif@cis.asu.edu.eg.}

\date{} %
\captionsetup{font={small}}
\maketitle
\begin{abstract} 
Vision Transformers (ViTs) have achieved significant advancement in computer vision tasks due to their powerful modeling capacity. However, their performance notably degrades when trained with insufficient data due to lack of inherent inductive biases. Distilling knowledge and inductive biases from a Convolutional Neural Network (CNN) teacher has emerged as an effective strategy for enhancing the generalization of ViTs on limited datasets. Previous approaches to Knowledge Distillation (KD) have pursued two primary paths: some focused solely on distilling the logit distribution from CNN teacher to ViT student, neglecting the rich semantic information present in intermediate features due to the structural differences between them. Others integrated feature distillation along with logit distillation, yet this introduced alignment operations that limits the amount of knowledge transferred due to mismatched architectures and increased the computational overhead. To this end, this paper presents \textbf{H}ybrid \textbf{D}ata-efficient \textbf{K}nowledge \textbf{D}istillation (HDKD) paradigm which employs a CNN teacher and a hybrid student. The choice of hybrid student serves two main aspects. First, it leverages the strengths of both convolutions and transformers while sharing the convolutional structure with the teacher model. Second, this shared structure enables the direct application of feature distillation without any information loss or additional computational overhead. Additionally, we propose an efficient light-weight convolutional block named \textbf{M}o\textbf{b}ile \textbf{C}hannel-\textbf{S}patial \textbf{A}ttention (MBCSA), which serves as the primary convolutional block in both teacher and student models. Extensive experiments on two medical public datasets showcase the superiority of HDKD over other state-of-the-art models and its computational efficiency. Source code at: 
\url{https://github.com/omarsherif200/HDKD}\\

\end{abstract}

\section{Introduction}
Accurate analysis of medical images is integral for successful treatment strategies. Over recent years, notable progress has been made in medical image analysis and classification, largely driven by the impressive progress of deep learning approaches. However, achieving accurate results with deep learning models requires extensive annotated data, which is considered a critical challenge, especially in the context of medical imaging. Knowledge transfer techniques \cite{zhuang_comprehensive_2021, gou_knowledge_2021} have been effective in addressing challenges associated with training on limited datasets. Among these techniques, transfer learning is widely adopted, which involves utilizing the knowledge gained from training a model on a source task, such as image classification on ImageNet \cite{deng_imagenet_2009}, and adapting it to a different target task, such as medical image classification. However, transfer learning comes with some limitations. Its effectiveness diminishes when tasks or datasets are significantly different. Additionally, making architectural adjustments is constrained by the fixed pre-defined parameters, leading to misalignment in weights during fine-tuning.
\newline
\indent Another popular technique is knowledge distillation \cite{hinton_distilling_2015}, where knowledge is transferred from a teacher model to a student model. This technique has been widely employed to enhance model performance and achieve model compression. Unlike transfer learning, knowledge distillation allows for greater flexibility in tuning the student architecture, making it a preferred choice in the realm of model compression \cite{polino_model_2018, li_few_2020, vedaldi_online_2020, avidan_tinyvit_2022}. Recently, knowledge distillation has been widely incorporated in vision transformers (ViTs) \cite{dosovitskiy_image_2020}, facilitating the transfer of inductive biases from a well-trained ConvNet teacher or ensemble of ConvNet teachers to a vision transformer model \cite{touvron_training_2021,ren_co-advise_2022}. However, this approach is effective when employed with logit-based knowledge distillation techniques, where knowledge is transferred from CNNs to ViTs by aligning their output layer probability distributions, but feature-based distillation techniques encounter challenges due to misalignment of intermediate feature representations between ConvNets and ViTs, resulting in information loss and suboptimal knowledge transfer.
\newline
\indent Accordingly, previous approaches \cite{avidan_tinyvit_2022,touvron_training_2021,ren_co-advise_2022, ahmadabadi_distilling_2023} to knowledge distillation have primarily focused on logit-based distillation between CNNs and ViTs. This method has been effective in improving the generalization and performance of ViTs. However, few studies have explored leveraging intermediate feature layers, which contain rich information but also present significant challenges due to the structural differences between CNNs and ViTs. CNNs are inherently designed to capture local patterns through inductive biases, whereas ViTs focus on capturing global dependencies across tokens. Consequently, the intermediate features captured by CNNs and ViTs differ in both semantic content and representational form, making feature-based distillation between them particularly challenging. Recent research \cite{chen_dearkd_2022, liu_cross-architecture_2024, avidan_locality_2022} has introduced alignment modules to address these challenges, however, these solutions often lead to information loss and increased computational overhead, as the rich, original features are transformed into a compressed form for alignment. Furthermore, they do not fully address the semantic differences between the two architectures, resulting in suboptimal knowledge transfer. 
\newline
\indent In this paper, we propose a simple yet effective design of a novel teacher-student paradigm called Hybrid Data-Efficient Knowledge Distillation Network (HDKD). HDKD seamlessly integrates both logit and feature-based distillation techniques to enable a more comprehensive knowledge transfer. HDKD paradigm strategically designs the teacher as a pure CNN and the student as a hybrid model to enable a shared convolutional structure between them. This design eliminates the need for additional alignment modules, preserving the richness of intermediate features. It also enables effective knowledge transfer by allowing distillation between features with consistent semantic information. We further introduce Mobile Channel-Spatial Attention (MBCSA) block as the primary convolutional block utilized in both teacher and student models. The student hybrid design serves two main purposes: it leverages the strengths of both convolutional inductive biases and the global processing capabilities of transformers, and it enables to apply feature distillation directly without any information loss and with minimal overhead, thanks to the shared convolutional structure between the proposed teacher and student networks.
\newline
\indent To the best of our knowledge, we are the first to investigate knowledge distillation between a CNN teacher and a hybrid student, eliminating the misalignment challenges inherent in applying feature-based methods between a CNN teacher and a ViT student. Furthermore, both teacher and student models are designed in a light-weight fashion, to improve the efficiency of the entire learning process and to ease the deployment of the student model on edge devices.
\newline
\indent We evaluate our models on two distinct medical datasets: the brain tumor dataset \cite{noauthor_brain_nodate} and the HAM-10000 (“Human Against Machine with 10000 training images”) dataset \cite{tschandl_ham10000_2018}. Our models outperformed other state-of-the-art models on both datasets, which demonstrates the robustness and adaptability of our methodology across various tasks and datasets. 
\\ \\ In summary, the contributions of this study are as follows:

\begin{enumerate}
    \item We present a novel teacher-student paradigm called Hybrid Data-Efficient Knowledge Distillation (HDKD). In particular, we propose a CNN teacher, and a hybrid student which alleviates the benefits of both CNNs (inductive biases) and Transformers (global processing capabilities). Moreover, we design the teacher and the student models to share a common structure in the convolutional stages, enabling the direct application of feature distillation from the teacher to the student.
    \item We introduce an efficient light-weight convolutional block named MBCSA block, which serves as the primary convolutional block in both teacher and student models.
    \item Unlike previous methods, HDKD is the first distillation paradigm to utilize a CNN teacher and a hybrid student. Additionally, it employs direct feature distillation, eliminating information loss caused by the alignment operations required in prior CNN-to-ViT distillation techniques.
    \item Extensive experiments validate that HDKD significantly enhances the performance, especially with limited data. Furthermore, HDKD outperforms state-of-the-art models while maintaining efficient and light-weight design, making it suitable for real-time deployment.
\end{enumerate}

\section{Related work}\label{sec2}
\subsection{Convolutional Neural Networks} CNNs have shown significant advancements in deep learning and have been the dominant networks especially for computer vision tasks. Over the last decade, various architectures have been introduced including AlexNet \cite{krizhevsky_imagenet_2012}, VGG networks \cite{simonyan_very_2015}, ResNets \cite{he_deep_2016}, DenseNets \cite{huang_densely_2017}, and Inceptions \cite{szegedy_going_2015}. Those architectures have emerged as dominant models, delivering state-of-the-art results. However, their high computational cost has limited their applicability to mobile and embedded devices. This is primarily attributed to the resource-intensive convolution operation of CNNs. To mitigate this limitation, another line of research developed efficient light-weight CNN models for mobile vision tasks with fewer computational parameters such as MobileNets \cite{howard_mobilenets_2017, sandler_mobilenetv2_2018, howard_searching_2019}, ShuffleNet \cite{zhang_shufflenet_2018}, EfficientNets \cite{tan_efficientnet_2019,tan_efficientnetv2_2021}, etc. For instance, MobileNet replaces the expensive standard convolution with an efficient separable depth-wise convolution. ShuffleNet focuses on reducing model parameters and computations through utilizing channel shuffling and pointwise group convolutions. EfficientNet utilizes a scaling technique that adjusts the network's resolution, width and depth, to achieve a good balance between performance and efficiency. 

\subsection{Transformers} Transformers was initially introduced in 2017 by Vaswani et al. \cite{vaswani_attention_2017} for machine translation tasks. Since then, it has been the dominant paradigm for natural language processing (NLP) tasks by enabling the efficient encoding of long-range dependencies through the self-attention mechanism. Recently, ViT \cite{dosovitskiy_image_2020} has extended the application of transformers to computer vision tasks, achieving state-of-the-art performance when pretrained on extensive datasets such as JFT-300M \cite{sun_revisiting_2017}. However, ViTs still lag behind CNNs when being trained on limited datasets because they lack inductive biases \cite{dascoli_convit_2021,li_localvit_2021,xu_vitae_2021}. Fortunately, ViTs can still achieve CNN-level performance when being trained on less compact datasets (e.g., ImageNet \cite{deng_imagenet_2009}) by incorporating different techniques such as applying knowledge distillation and extensive data augmentation techniques \cite{touvron_training_2021}. Beyond ViTs, other studies have explored different transformer architectures that inherit the structure of CNN models by reducing the spatial dimension and increasing the depth as the network progresses, making them suitable for down streaming tasks \cite{liu_swin_2021,liu_swin_2022,wang_pyramid_2021}. Another line of research focused on reducing the complexity of self-attention mechanism by introducing efficient local attentions \cite{liu_swin_2021,liu_swin_2022,chu_twins_2021, chen_regionvit_2021}. 

\subsection{Knowledge distillation} Knowledge distillation \cite{hinton_distilling_2015} was originally proposed in 2015 by Geoffrey Hinton et al. The aim of proposing this method is to transfer knowledge from a larger, more complex teacher model to a smaller, more compact student model to match certain outputs. Knowledge distillation has gained significant attention in deep learning for its efficiency in various tasks such as model compression \cite{polino_model_2018,li_few_2020,vedaldi_online_2020}, domain adaption \cite{kothandaraman_domain_2021, reddy_domain-aware_2024}, model specialization on small datasets \cite{touvron_training_2021, liu_network_2023}, etc. Numerous studies have leveraged the benefits of knowledge distillation. Some have centered their focus on logit-based methods \cite{hinton_distilling_2015, touvron_training_2021, ren_co-advise_2022, huang_knowledge_2022} to enhance the student predictions, while others focused on feature-based methods \cite{avidan_locality_2022, heo_comprehensive_2019, zhang_task-oriented_2020, park_relational_2019} to enrich the student learning representations of intermediate features, with guidance from the teacher model. Knowledge distillation success extends beyond CNNs, showing remarkable efficiency in vision transformers. For example, DeiT \cite{touvron_training_2021} improves the data efficiency by distilling the knowledge from CNNs to vision transformers through a distillation token. CiT \cite{ren_co-advise_2022} guides the student transformer network to distill knowledge from light-weight teachers with different inductive biases. DearKD \cite{chen_dearkd_2022} follows a two-step process. It starts by distilling inductive biases from early intermediate CNN layers, followed by allowing the transformer to take full control and train without distillation. CSKD \cite{zhao_cumulative_2023} transfers spatial knowledge directly from CNNs spatial responses to all corresponding patch tokens of ViT, without introducing intermediate features.

\subsection{Hybrid Models} Although pure vision transformers have achieved state-of-the-art results by effectively capturing long-term dependencies through self-attention mechanism. They do come with certain drawbacks. For instance, these models are difficult to train and require high computational resources due to their reliance on self-attention. In addition, they lack the inductive biases inherent in CNNs and demand larger amounts of data and extensive augmentation techniques to achieve comparable performance to CNNs, which excel at generalizing with limited data. From this pursuit, researchers have explored various hybrid designs that combine both transformers and convolutions \cite{dai_coatnet_2021, avidan_maxvit_2022, yang_moat_2022, wang_pvt_2022}, aiming to leverage the benefits of both approaches. Additionally, there have been efforts to create light-weight hybrid designs suitable for deployment on resource-constrained devices. Examples include Mobile-Former \cite{chen_mobile-former_2022} which introduces a parallel structure of MobileNet and transformers, where both communicate through a bidirectional bridge using a cross-attention mechanism. MobileViT \cite{mehta_mobilevit_2021} views transformers as convolutions, allowing for the combination of both in an efficient block that captures global representations. MobileViTv2 \cite{mehta_separable_2022} builds upon MobileViT, but replaces the bottleneck multi-headed-self-attention (MHSA) in transformers with a separable self-attention layer which operates with linear complexity. 

\section{Methodology}\label{sec3}
In this section, we briefly provide an overview of the mobile inverted bottleneck convolution (MBConv) block with the squeeze-and-excitation (SE) module in addition to the transformer block. Subsequently, we delve into our proposed teacher-student paradigm (HDKD) and the MBCSA block.

\subsection{Overview of Light-weight convolution block (MBConv)}
MBConv \cite{sandler_mobilenetv2_2018}, known as inverted residual block, was introduced in 2018. Since then, it was widely utilized in different mobile vision tasks due to its efficiency compared to the standard expensive convolutional layers. The MBConv block (see Figure 1) is structured as follows: a 1\(\times\)1 convolutional layer (expansion layer) is employed to expand the input channels by a factor of 4, followed by a 3\(\times\)3 depthwise convolutional layer to encode the spatial local interactions followed by the squeeze-and-excitation (SE) module \cite{hu_squeeze-and-excitation_2018}, which learns channel-wise feature dependencies and dynamically adjusts the relevance of each channel, ensuring that the network focuses on the most informative features. Finally, a 1\(\times\)1 convolutional layer (projection layer) is used to shrink back the feature channels into the original dimension, enabling a residual connection with the input image. Due to its efficiency and lightweight design, MBConv is widely incorporated in the state-of-the-art CNN architectures \cite{sandler_mobilenetv2_2018, tan_efficientnet_2019, tan_efficientnetv2_2021, tan_mnasnet_2019}. Moreover, recent state-of-the-art hybrid architectures that seek efficiency with reasonable computational cost have adopted the MBConv block as their primary convolutional unit \cite{dai_coatnet_2021, chen_mobile-former_2022, mehta_mobilevit_2021}.

\subsection{Overview of Transformer Block}
The Transformer block \cite{vaswani_attention_2017} effectively captures global dependencies present within the data through its self-attention operation. It comprises two core components, namely Multi-headed self-attention (MHSA) and Multi-layer perceptron (MLP). MHSA enhances the model’s capability to prioritize significant regions in the image. In addition, it affords a broader contextual understanding compared to traditional convolutional layers. The self-attention computes the pairwise similarity between each patch (group of pixels) in the input tensor, enabling the model’s receptive field to cover the entire spatial domain. The attention mechanism module is defined as follows: 
\begin{equation} 
\label{eq:Attention}
    \text{Attention}(Q, K, V) = \text{softmax}\left(\frac{QK^T}{\sqrt{d_k}}\right)V
\end{equation}
The input tensor is partitioned into \(n\) equal-sized patches \(X \in \mathbb{R}^{n \times d}\) and is linearly transformed into 3 vectors, i.e., Query Vector \((Q \in \mathbb{R}^{n \times d_k})\), Key Vector \((K \in \mathbb{R}^{n \times d_k})\), and the Value Vector \((V \in \mathbb{R}^{n \times d_v})\), where \(Q = W_Q . X\), \(K = W_K . X\), and \(V = W_V . X\), where \(W_Q\), \(W_K\), and \(W_V\) are the projection learnable weight matrices. The scaled dot product is computed based on these three vectors (Q, K, and V). The MLP layer consists of two 1\(\times\)1 convolutional layers. In the first layer, the channels are expanded by a factor of 4. In the second one, they are projected back and GeLU non linearity activation function \cite{hendrycks_gaussian_2023} is applied between them.

\begin{equation} 
\label{eq:MLP}
    \text{MLP}(X) = \text{Conv}(\text{GeLU}\text(\text{Conv}(LN\text(X))))
\end{equation}
Where LN refers to Layer Normalization \cite{ba_layer_2016}.

\subsection{MBCSA Block}
The proposed MBCSA (\textbf{M}o\textbf{b}ile \textbf{C}hannel-\textbf{S}patial \textbf{A}ttention) block is a pure convolutional block that seamlessly integrates the MobileNet block with the CBAM module \cite{ferrari_cbam_2018}. MBCSA block enhances over MBConv block by incorporating CBAM module \cite{ferrari_cbam_2018} instead of SE module \cite{hu_squeeze-and-excitation_2018}. This modification allows the MBCSA block to incorporate robust integration of channel and spatial attention mechanisms, providing a more comprehensive feature refinement compared to the original MBConv block, which only employs channel attention via the SE module. The channel attention (CA) mechanism enables the model to focus on the most informative features by computing a set of attention weights for each channel and dynamically adjusting the relevance of each channel. The spatial attention (SA) mechanism computes a set of attention weights for each spatial location in the feature map, allowing the model to prioritize relevant spatial regions. In CBAM, both attention mechanisms are computed through convolution rather than using transformer-based mechanisms. In addition, the ReLU activation function in the channel attention mechanism of the CBAM block is replaced with GELU activation \cite{hendrycks_gaussian_2023}, for enhancing the features adaptability by introducing a richer set of non-linearities. The CBAM module is given as follows:

\begin{equation} 
\label{eq:CBAM}
    F' = M_S \left( M_C (F) \otimes F \right) \otimes F
\end{equation}

Where \(F\) is the input feature map, \(F \in \mathbb{R}^{H \times W \times C}\). \(M_C\) is the channel attention mechanism where \(M_C (F) \in \mathbb{R}^{1 \times 1 \times C}\). \(M_S\) is the spatial attention mechanism where \(M_S (F) \in \mathbb{R}^{H \times W \times 1}\), and \(\otimes\) is an element-wise multiplication operation. During this multiplication, the channel attention values and the spatial attention values are broadcasted across the spatial dimension and the channels dimension, respectively.
\vspace{0.1cm}
\newline
The channel attention (CA) mechanism is given as follows:

\begin{equation}
\label{eq:CA}
\begin{aligned}
M_C (F) &= \sigma(\text{MLP}(\text{AvgPool}(F)) + \text{MLP}(\text{MaxPool}(F))) \\
\text{MLP}(x) &= \text{Linear}_{\text{expand}} (\text{GELU}(\text{Linear}_{\text{shrink}}(x)))
\end{aligned}
\end{equation}

The given feature is aggregated using max-pooling and average pooling operations along the spatial dimensions. Subsequently, both aggregated results are passed through a shared multi-layer perceptron (MLP), and their outputs are added together. Lastly, a sigmoid function \((\sigma)\) is applied to produce the probabilistic channel attention values (\(M_C (F) \in \mathbb{R}^{1 \times 1 \times C}\)). The MLP layer comprises two linear layers with a GELU activation in between. The first linear layer shrinks the input with a factor \(r\) while the second linear layer expands the input back.
\vspace{0.1cm}
\newline
The spatial attention (SA) mechanism is given as follows:
\begin{equation} 
\label{eq:SA}
    M_S (F) = \sigma(\text{Conv}_{7\times7} (\text{Concat}(\text{AvgPool}(F), \text{MaxPool}(F))))
\end{equation}

The given feature is aggregated using max-pooling and average pooling operations along the channel dimensions and their outputs are concatenated to produce an output of size (\(H\times W\times2\)). The resulting output is passed to a convolutional layer with only one filter of size \(7\times7 \) to yield an output of size (\(H \times W \times 1\)). Lastly, a sigmoid function \((\sigma)\) is applied to produce the probabilistic spatial attention values (\(M_S (F) \in \mathbb{R}^{H \times W \times 1}\)). The MBConv block and the proposed MBCSA Block are depicted in Figure 1.

\begin{figure}[tb!]
\centering
    \includegraphics[width=0.98\linewidth]{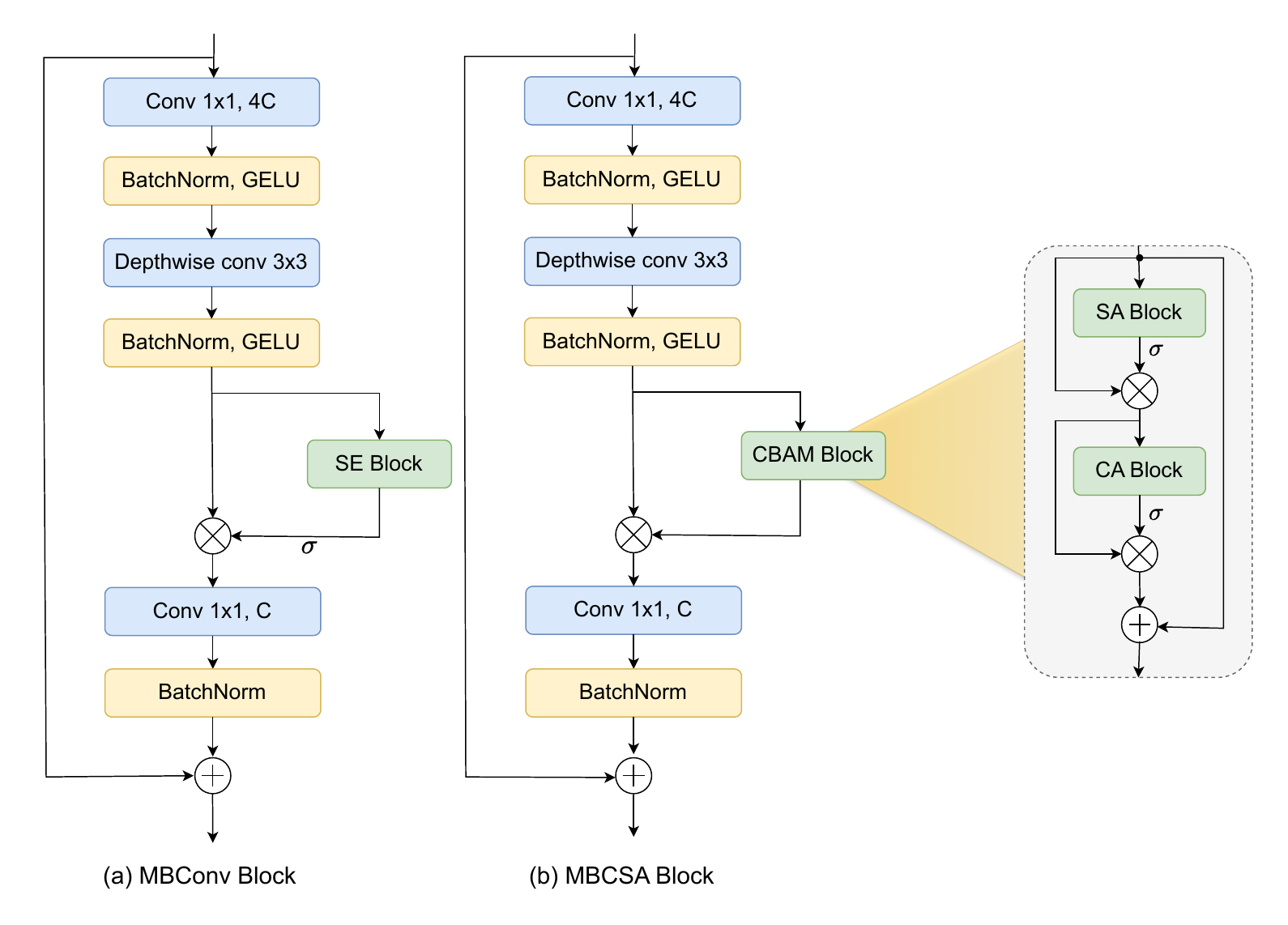}
    \caption{\textbf{Comparison between MBConv block and MBCSA Block.} (a) is the standard MobileNet block with SE module, (b) is the proposed MBCSA block which replaces the SE module with CBAM module to adjust the relevance of both spatial and channel information.}
\end{figure}

\subsection{Hybrid Data-Efficient Knowledge Distillation (HDKD) Network}
In this study, we present HDKD, a novel light-weight architecture based on a teacher-student paradigm as represented in Figure 2. The proposed approach combines the strengths of both convolutional neural networks (CNNs) and vision transformers (ViTs) by employing a hybrid model as the student network. It leverages the benefits of using distillation which aims to make the student model generalize well, even when trained with limited data and samples \cite{li_few_2020, avidan_black-box_2022, cui_kd-dlgan_2023}. The teacher network follows a pure CNN model and consists of three stages followed by pooling and fully connected layers. The student network consists of four stages, sharing the same hierarchy as the teacher network for the first three stages, while the last stage employs a Distilled Feature-level Transformer (DFLT) module. These shared stages include a stem block in the first stage and MBCSA blocks in the second and third stages. Each stage reduces the input spatial dimension with a factor of 2 and increases the channel dimensions. The stem block consists of a convolutional layer with a 3\(\times\)3 kernel, followed by batch normalization and GeLU activation function. Unlike recent studies \cite{chen_dearkd_2022, liu_cross-architecture_2024, avidan_locality_2022}, our proposed paradigm allows direct feature distillation between the student and teacher models without additional alignment operations, thanks to the shared hierarchy in the first three stages. This shared hierarchy also avoids any potential loss of feature information during distillation. 
\newline
\indent We adopt a strategic design choice by incorporating a larger number of blocks per stage in the teacher network compared to the corresponding stages in the student network. Specifically, \(N_i > M_i \quad \forall  i \), where \( N_i \) and \( M_i \) are the number of blocks at stage \( i \) for the teacher and student models, respectively. This design choice serves two main purposes. Firstly, it empowers the teacher model to acquire enhanced feature representations, which can then be effectively transmitted to the student model through knowledge distillation. Secondly, this design helps to reduce the size and computational demands of the student model, making it lighter and more computationally efficient.
\newline
\indent The final stage of the student network, which employs the DFLT module, is built upon the design of DeiT network \cite{touvron_training_2021}, with a notable modification. Instead of taking the original raw image as input, the DFLT module takes high-level features obtained from the convolutional layers in the first three stages. This allows the transformer block to process global information present in the feature-level representation, rather than operating at the pixel-level representation. By leveraging this feature-level information, the DFLT module can capture richer contextual information and improve the overall performance of the student model.  Additionally, the inclusion of convolutional operations before vision transformers introduces the inductive biases to the network and reduces the complexity required for self-attention layer computations. Both the CLS and Distillation tokens in DFLT module serve as global token representations, interacting with other tokens through attention. Additionally, they are optimized to match the actual label and teacher logits, respectively.
\newline
\indent The training paradigm of the HDKD network consists of two steps. In the first step, the teacher model is trained on the entire data with robust augmentations to learn generalized representations and patterns. In the second step, the student model is trained across different data sizes, with knowledge transferred from the teacher model through feature and logit distillation. Training across different data sizes is crucial to demonstrate the model's ability to generalize well, even when trained with limited data. During inference, predictions are made using the student model by averaging the class (CLS) logits and distillation logits. 

\begin{figure}[tb!]
\centering
    \includegraphics[width=\linewidth]{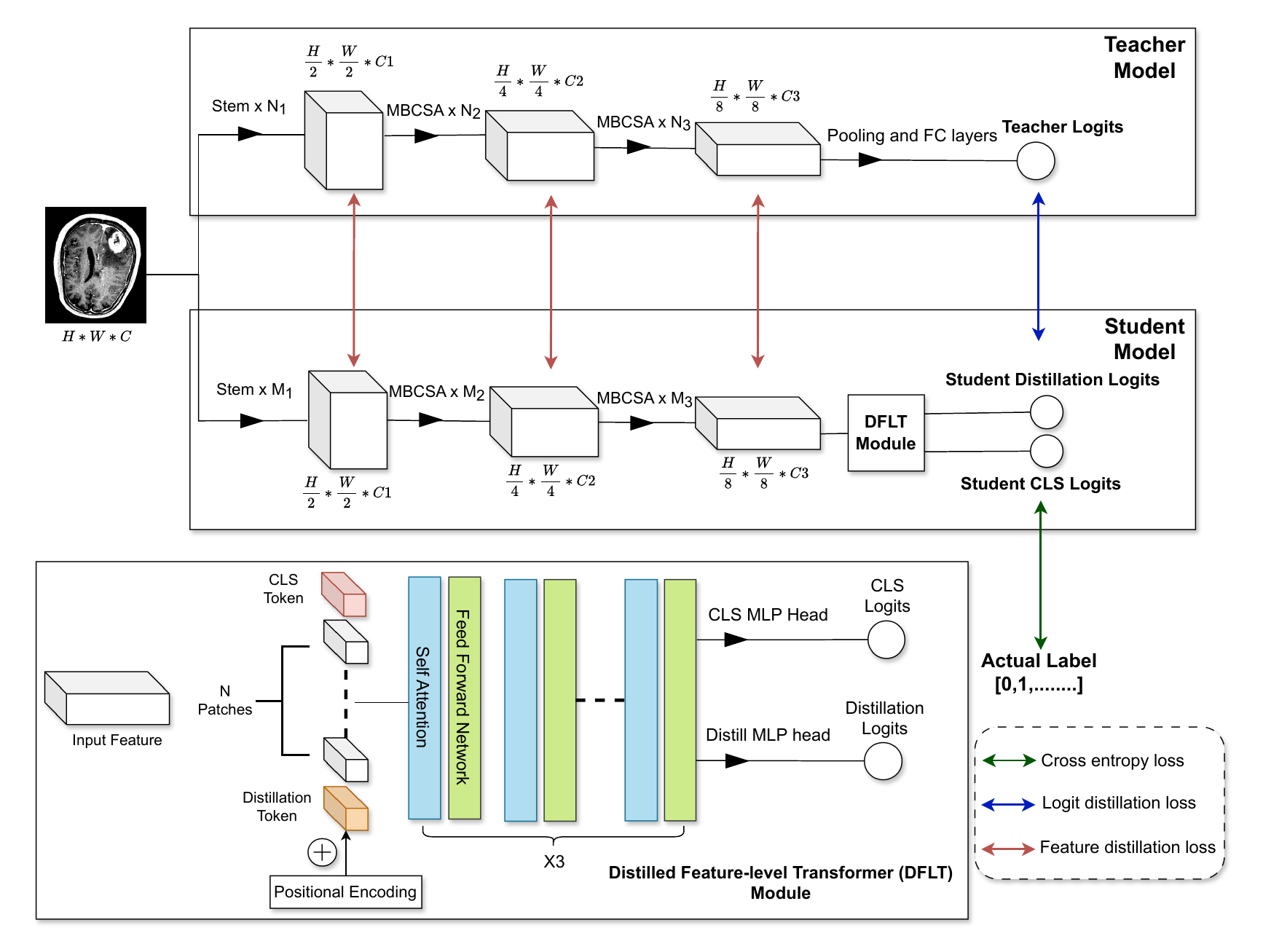}
    \caption{\textbf{Overview of the proposed teacher-student (HDKD) paradigm.} The process involves initially training the teacher model, followed by training the student model. During training the student model, it leverages the knowledge distilled from the teacher model through logit and feature distillation techniques.}
\end{figure}

\subsection{Distillation techniques}
In this section, we delve into the distillation techniques and loss functions employed to optimize the proposed models. While traditional classification neural networks commonly utilize cross-entropy loss to align model predictions with actual labels, teacher-student paradigm models go beyond mere label approximation. These paradigms also aim to mimic the distribution between the student model’s predictions and the actual labels, as well as the distribution between the student and teacher predictions. This is achieved through a technique called logit distillation which tries to transfer knowledge from the teacher model to the student model by aligning their logit distributions. This allows the student model to capture new relevant information from the teacher model, such as understanding object visibility in an image or the effects of augmentations on altering the original image \cite{wei_circumventing_2020}. Such additional information goes beyond what can be inferred from the actual labels alone. For instance \cite{hinton_distilling_2015,touvron_training_2021,wei_circumventing_2020} uses Kullback-Leibler divergence (KL) loss in logit distillation to approximate the student predictions towards the teacher predictions. In our proposed paradigm, the teacher model undergoes initial training using cross-entropy (CE) loss only. Meanwhile, the optimization of the HDKD model's classification loss involves a weighted combination of cross-entropy (CE) loss and Kullback-Leibler divergence (KL) loss, where KL loss is responsible for applying logit distillation. The classification objective \(( l_{cls} \)) of our HDKD model is given as follows:

\begin{equation} 
\label{eq:LKD}
    l_{\text{cls}} = (1 - \alpha) \, l_{\text{CE}}(\text{softmax}(z_s), y_a) + \alpha \, t^2 \, l_{\text{KL}}(\text{softmax}(\frac{z_s}{t}), \text{softmax}(\frac{z_t}{t}))  
\end{equation}

The cross-entropy loss (\( l_{CE} \)) is used as the first term to align the student’s prediction with the actual label, while the Kullback-Leibler divergence (\( l_{KL} \)) is used as the second term to align the student’s prediction with the teacher’s prediction, Let \( Z_s \) and \( Z_t \) be the logits of the student model and the teacher model, respectively.  \( \alpha \)  is a hyperparameter to balance the trade-off between the two sub-losses while \( t \) refers to temperature and it controls the softness of the probability distribution. The SoftMax function is used to convert the given logits into probabilities. We set \( \alpha \) and \( t \) in all our experiments to 0.5 and 1, respectively. 
\newline
\indent Although logit distillation effectively transfers knowledge, its notable limitation lies in its sole focus on the final prediction layer. This can limit the comprehensive transfer of knowledge, especially regarding the features information captured within the intermediate layers. To address this limitation, we incorporate feature distillation into our proposed paradigm. The feature distillation step involves aligning the features representations in the intermediate layers to minimize the disparity between them. By doing so, the student model will not only benefit from approximating the teacher’s final prediction but also gain complex insights and relationships captured from the deeper hierarchical features learned by the teacher model. Feature distillation is applied directly between the features in the intermediate layers without any alignment operations due to the shared hierarchy, as stated in section 3.4.  
\newline
\indent Given the teacher model's accessibility to more data compared to the student model in addition to its larger number of blocks per convolutional stage. This direct feature-to-feature mapping enriches the student model, facilitating the acquisition of generalized patterns from the teacher model and mitigating the risk of overfitting. We employ feature distillation in the three shared stages (see Figure 2), with higher weights assigned to later stages that contain high-level features, since high-level features are more task-specific and carry more semantic information compared to low-level features. 
\newline
\indent Feature distillation also helps the DFLT module to receive generalized features coming from the teacher model as an input, which enhances the transformer's capability to capture more robust semantic features, thereby improving the overall global processing of the network. The feature distillation objective (\( l_{feat} \)) of our HDKD model is defined as follow:

\begin{equation} 
\label{eq:FKD}
    l_{\text{feat}} = \sum_{j=1}^{N-1} \frac{1}{H_j W_j C_j} \| \psi_j^T - \psi_j^S \|_2^2 \times j^\alpha 
\end{equation}

Where \( \psi_j^T \) and \( \psi_j^S \) represent the feature maps obtained from the last MBCSA block at a specific stage (stage \( j \)) of the teacher and student models, respectively, and \( j \in \{1, \ldots, N\} \), where \( N \) is the number of stages. \( H_j \), \( W_j \), and \( C_j \) represent the height, width, and number of channels of the feature maps at stage \( j \). Each feature loss component at each stage is computed using mean-squared error (MSE) loss. The multiplication part (\( \times j^\alpha \)) is responsible for weighting each feature loss component, so higher weights are given to later features as those features are more task-oriented and carry more semantic information. \( \alpha \) is set to 1 in all the experiments. The overall loss for optimizing the HDKD network combines the classification loss and the feature distillation loss and is defined as follows:

\begin{equation} 
\label{eq:total loss}
l_{\text{combined}} = l_{\text{cls}} + \lambda \cdot l_{\text{feat}}
\end{equation}

Here, \( \lambda \) serves as a regularization hyperparameter, controlling the impact of the feature distillation loss (\( l_{\text{feat}} \)) on the overall optimization loss. It allows to control the trade-off between the classification loss (\( l_{\text{cls}} \)) and the feature distillation loss (\( l_{\text{feat}} \)) during the training process. In our experiments, we fix \( \lambda \) to 10.

\section{Experiments \& Results}\label{sec4}
In this section, we evaluate our proposed models on diverse medical benchmark datasets. Additionally, we present detailed experiments and analysis of the obtained results.
\subsection{Dataset}
In this study, the proposed models are evaluated on two different medical datasets for two distinct diseases; brain tumor and skin cancer. The first dataset (Brain Tumor MRI dataset) \cite{noauthor_brain_nodate} contains data from three benchmark databases including Figshare, SARTAJ, and Br35H. It consists of four classes (i.e., no tumor, pituitary, meningioma, and glioma) with total number of 7023 images. Each class consists of more than 1600 samples. The second dataset used is HAM-10000 \cite{tschandl_ham10000_2018}, a large benchmark dataset featuring 10015 dermatoscopic images, each sized at 450\(\times\)600 pixels. This dataset consists of seven categories of skin lesions: Dermatofibroma (DF), Intra-Epithelial Carcinoma (AKIEC), Benign Keratosis (BKL), Melanoma (MEL), Vascular Lesions (VASC), Melanocytic Nevi (NV) and Basal Cell Carcinoma (BCC). The dataset represents a critical challenge due to its highly imbalanced class distribution, ranging from 6705 images for the largest class (NV) to 115 for the smallest class (DF), where the NV class alone represented more than 60\% of the data. In the brain tumor dataset, the training and testing sets consist of 5711 and 1311 samples, respectively. Meanwhile, in HAM-10000 dataset, we follow the same dataset division used in \cite{datta_soft-attention_2021}, with 9187 samples in the training set and 828 samples in the testing set.

\subsection{Implementation details}
The proposed models are trained from scratch on two different medical datasets, employing an image resolution of 224\(\times\)224 for both the training and testing sets of each dataset. All the experiments are carried out using PyTorch on GPU P100, with a batch size of 32. The teacher model is trained using Adam optimizer \cite{kingma_adam_2015} with a fixed learning rate of 0.001 for both datasets. while the student model is trained using AdamW optimizer \cite{loshchilov_decoupled_2019} with a cosine learning rate scheduler, an initial learning rate of 5e-5, a minimum learning rate of 2e-5, a linear warmup for 5 epochs and a weight decay of 0.1. During the training process, the teacher model relies on extensive data augmentation and balancing techniques (i.e., basic data augmentation techniques and smote \cite{chawla_smote_2002}). These techniques play a crucial role in the teacher model's comprehensive understanding of diverse patterns in the data. Subsequently, through the distillation process, this enriched knowledge gained from the teacher is effectively transferred to enhance the learning process of the student model and its adaptability to unseen variations.

\subsection{Architectures specifications}
The architecture specifications for both the proposed student and teacher models are given in Table 1. The number of MBCSA blocks and channels per stage for the teacher and student models are reported. Similarly, details regarding the DFLT module in the student model, including the number of transformer layers, patch size, embedding dimension, and number of heads along with the dimension per head are reported. Furthermore, we state the number of parameters and FLOPs for each model. Notably, the teacher model incorporates a larger number of convolutional blocks per stage compared to the student model. This design choice not only enables the teacher model to capture more intricate details and facilitate effective knowledge transfer to the student model, but also contributes to the compression of the student model by reducing its overall number of blocks. As a result, the student model has 2.08G FLOPs which is significantly lower than the teacher model's 5.88G FLOPs, due to processing feature maps with fewer blocks.
 
\renewcommand{\arraystretch}{1.2}
\begin{table*}[h]
\centering
\caption{Teacher and student architecture specifications.}
\small 
\begin{tabular}{ p{2.1cm}|p{1.3cm} p{1.6cm}|p{1.75cm} p{0.8cm} p{1.5cm} p{0.88cm}|p{1.1cm} p{0.8cm}  }
 \hline
 \empty & \multicolumn{2}{c|}{Convoluion part} & \multicolumn{4}{c|}{Transformer Part (DFLT Module)} & \multicolumn{2}{c}{ } \\ 
 \hline
 Model & \#MBCSA per stage & \#Channels per stage & \#Transformer layers & Patch size & Embedding dimension & Heads x Dim & \#Params & FLOPs\\ 
 \hline
 Teacher model & [6,6,9] & [64,128,192] & \multicolumn{4}{c|}{N/A} & 3.95M & 5.88G\\
 Student model & [2,2,3] & [64,128,192] & 3 & (2,2) & 256 & 8 x 32 & 3.72M & 2.08G\\ 

 \hline
\end{tabular}
\end{table*}
 
In the following sections, we compare the student model with distillation (HDKD) against the student model without distillation. This comparison is necessary to observe the generalization capability of HDKD over its non-distilled counterpart. In HDKD, the teacher model guides the student model through logit and feature distillation during the training process. It incorporates a distillation token in the DFLT module and is optimized using a combined loss function (see Equation 8) to effectively learn generalized distributions and patterns. In contrast, the student model only (i.e., without distillation) doesn’t incorporate a distillation token and is optimized using cross entropy (CE) loss only, without any guidance from the teacher model. Both the distilled and non-distilled versions are trained using AdamW optimizer with the same settings mentioned in Section 4.2. Table 2 provides a detailed comparison between HDKD and the non-distilled student model, showing that the parameters and FLOPs are nearly identical, with a slight increase in HDKD due to the inclusion of the distillation token. 
\begin{table*}[tb!]
\centering
\caption{Comparison between the HDKD and the student model without distillation.}
\begin{tabular}{lccccc}
\toprule
\multicolumn{1}{l}{Model} & \multicolumn{1}{c}{CLS token} & \multicolumn{1}{c}{Distill token} & \multicolumn{1}{c}{Loss function} & \multicolumn{1}{c}{\#Params} & \multicolumn{1}{c}{FLOPs} \\ \toprule
Student (w/o) distillation & \ding{51} & \ding{55} & $l_{CE}$  & 3.72M & 2.08G \\
HDKD & \ding{51} & \ding{51} & $l_{cls} + \lambda \cdot l_{feat}$ & 3.73M & 2.09G \\
\bottomrule
\end{tabular}
\end{table*}

\subsection{Analysis and Results}
This section presents a comprehensive analysis of the performance of our proposed teacher-student paradigm (HDKD) across both datasets. We illustrate how our paradigm, coupled with distillation techniques, enhances the generalization across varying data sizes. Furthermore, we conduct a comparative analysis against state-of-the-art models on both datasets. It is important to note that all models undergo training from scratch without pre-training, ensuring a fair comparison with the results obtained from our proposed models.
\subsubsection{Results using Brain Tumor dataset}
In the initial training phase, the teacher model undergoes training using the entire dataset with robust data augmentation techniques. It achieved a remarkable accuracy of 99.77\% when trained on the entire dataset. Subsequently, the student model undergoes training with varying data sizes, integrating both logit and feature distillation to enhance its generalization and effectively inherit comprehensive insights from the trained teacher model. Additionally, the student model undergoes training across diverse data sizes without distillation, to provide a comparison with its distilled counterpart (HDKD). Detailed results for the student model and its distilled version across different data sizes are reported in Table 3. We employ four distinct data sizes: 200 images, 400 images, 600 images, and the entire dataset, with the first three maintaining a balanced distribution across classes. Specifically, these first three sizes include a balanced selection of 50, 100, and 150 images per class, respectively. It can be observed from Table 3 that the student distilled version (HDKD) surpasses the non-distilled version with an absolute gain of 16.4\%, 13.12\%, 11.67\% and 0.08\% on the four data sizes, respectively. 
\begin{table*}[!b]
\centering
\caption{Comparison of our HDKD model with the student model only (i.e., without distillation) on Brain Tumor dataset. HDKD achieves favorable performance across all data sizes.}
\begin{tabular}{ p{4.2cm} p{2.2cm} p{2.2cm} p{2.2cm} p{2.2cm}  }
 \toprule
 \empty & 200 images & 400 images & 600 images & All data \\ 
 \toprule
 Student w/o distillation & 76.5\% & 82.0\% & 85.28\% & 99.77\% \\ 
 \textbf{HDKD} & \textbf{92.9\%} & \textbf{95.12\%} & \textbf{96.95\%} & \textbf{99.85\%} \\

 \bottomrule
\end{tabular}
\end{table*}

\subsubsection{Results using HAM-10000 dataset}
The training procedure employed for the Brain Tumor dataset is replicated for the HAM-10000 dataset. However, to address the highly imbalanced nature of the data, SMOTE \cite{chawla_smote_2002} is utilized in addition to the data augmentation techniques for training the teacher model. The teacher model achieved an accuracy of 91.79\% on the testing set. Table 4 shows the results of the student model and its corresponding distilled version (HDKD) across different data sizes. Similar to the brain tumor dataset, we employ here four data sizes: 350 images (i.e., 50 images per class), 700 images (i.e., 100 images per class), 991 images (i.e., 150 images per class except for DF and VASC classes as they contain less than 150 images) and the entire dataset. It can be observed from Table 4 that the student distilled version (HDKD) surpasses the non-distilled version with an absolute gain of 4.71\%, 4.11\%, 4.23\%, and 2.54\% on the four data sizes, respectively. 
\begin{table*}[htbp]
\centering
\caption{Comparison of our HDKD model with the student model only (i.e., without distillation) on HAM-10000 dataset. HDKD achieves favorable performance across all data sizes.}
\begin{tabular}{ p{4.2cm} p{2.2cm} p{2.2cm} p{2.2cm} p{2.2cm}  }
\toprule
 \empty & 350 images & 700 images & 991 images & All data \\ 
\toprule
 Student w/o distillation & 75.0\% & 77.29\% & 78.62\% & 89.61\% \\ 
 \textbf{HDKD} & \textbf{79.71\%} & \textbf{81.4\%} & \textbf{82.85\%} & \textbf{92.15\%} \\

\bottomrule
\end{tabular}
\end{table*}
\subsubsection{The Impact of Dataset Size}
In this section, we investigate how the dataset size influences the student model performance, and how the distillation strengthens its generalization, particularly when trained with limited data. Figure 3 and Figure 4 show the accuracies of the student model across different data sizes for the Brain Tumor MRI dataset and HAM-10000 dataset respectively, both with and without incorporating distillation. It can be observed that incorporating distillation leads to significant improvements in performance when keeping the data size constant. Furthermore, the performance gap between the distilled and non-distilled model narrows as the dataset size increases. This is because a larger and more diverse dataset allows the non-distilled student model to better understand the data distribution, reducing overfitting and thus narrowing the performance gap with the distilled model, which benefits from the generalized distributions captured from the teacher model.

\begin{figure}[tb!]
    \centering
    \begin{minipage}[b]{0.48\linewidth}
        \includegraphics[width=\linewidth]{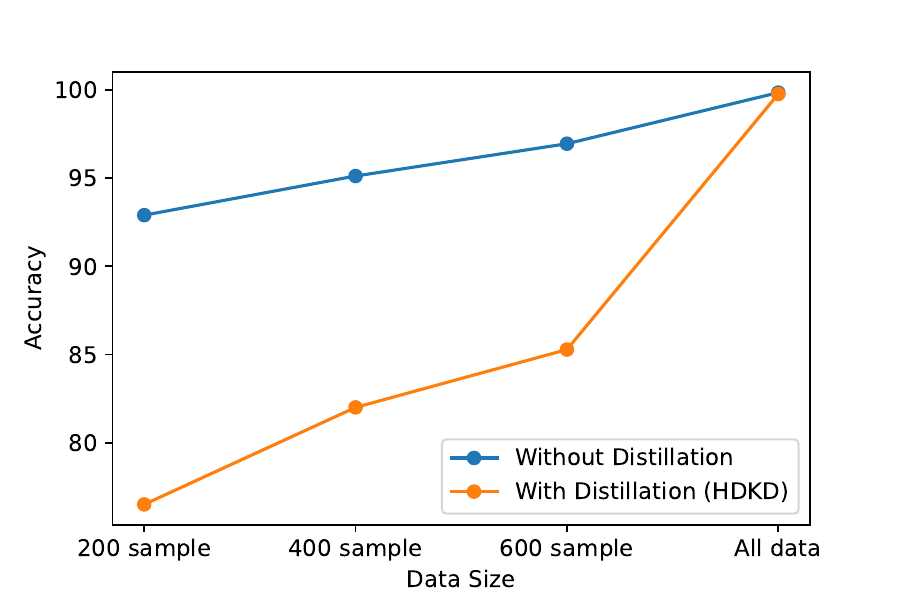}
        \caption{Comparison of the distilled student version (HDKD) with its non-distilled version across various data sizes on brain tumor MRI dataset.}
    \end{minipage}
    \hfill
    \begin{minipage}[b]{0.48\linewidth}
        \raisebox{88pt}{ 
        \parbox{\linewidth}{ 
                \includegraphics[width=\linewidth]{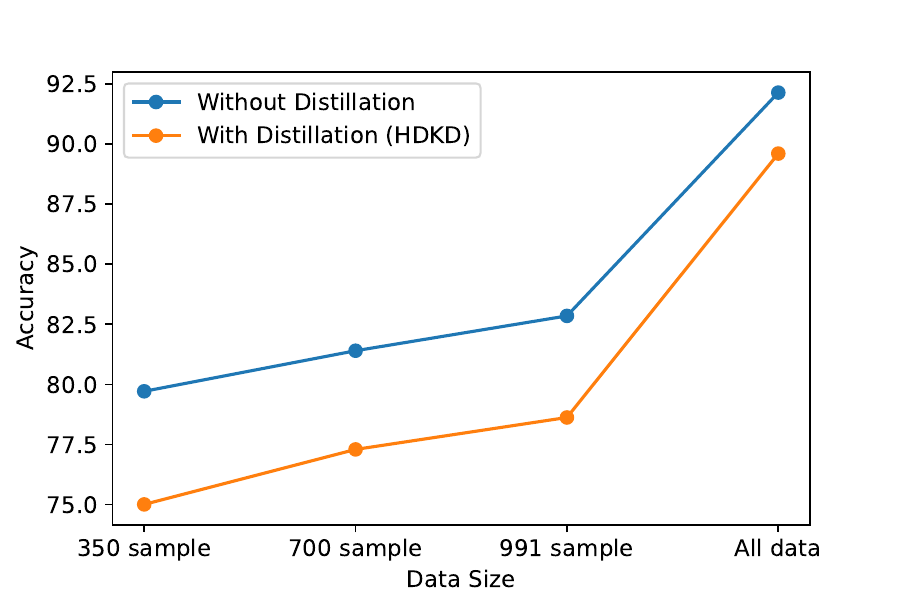}
                \caption{Comparison of the distilled student version (HDKD) with its non-distilled version across various data sizes on HAM-10000 dataset.}
            }
        }

    \end{minipage}
\end{figure}

\subsubsection{Comparison with state-of-the-art models}
In this section, we compare the proposed teacher, student and HDKD models with recent state-of-the-art light-weight models that have comparable computational costs and parameters. These models fall into three categories: Convolutional Neural Networks (ConvNets), Vision Transformers (ViTs), and Hybrid models. To ensure a fair comparison, all models undergo training from scratch, similar to our proposed ones. Table 5 presents an overview of the parameters, FLOPs, and accuracy achieved by these models across the two datasets. Notably, our HDKD model achieves superior accuracy on both datasets compared to other models, while also being cost effective, making it suitable for real-time applications.
\newline
\indent We also considered CSKD model \cite{zhao_cumulative_2023} in our comparison, a recent CNN-to-ViT distillation paradigm that utilizes RegNet \cite{radosavovic_designing_2020} as its teacher model and ViT as its student model. The authors of CSKD conducted several experiments and reported that it outperforms other state-of-the-art distillation methods, such as Deit \cite{touvron_training_2021} and DearKD \cite{chen_dearkd_2022}, across various datasets. In our comparison in Table 5, CSKD showed better generalization compared to other ViT models due to distillation, but our HDKD surpasses it with 2.29\% and 0.61\% on HAM-10000 and Brain tumor, respectively. This further demonstrates the superiority of our HDKD model in enhancing model generalization and performance through our efficient proposed distillation paradigm.
\newline
\indent Additionally, we assess our teacher, student, and HDKD models using other evaluation metrics, such as Precision, Recall, and AUC score, as shown in Tables 6 and 7, to provide a more comprehensive evaluation of the model’s performance. Since CoatNet-0 \cite{dai_coatnet_2021} and EfficientNetB0 \cite{tan_efficientnet_2019} outperformed other experimented models, as indicated in Table 5, we also compare them with our proposed HDKD model in terms of Precision, Recall, and AUC scores on both datasets, as presented in Tables 8 and 9.

\begin{table*}[!tb]
\centering
\caption{State-of-the-art comparison on HAM-10000 and Brain Tumor datasets. We compare our teacher model, student model (i.e., without distillation) and HDKD model with state-of-the-art models across both datasets.}
\begin{tabular}{ p{2.7cm} p{4.8cm} p{1.2cm} p{1.2cm} p{2cm} p{2cm}}
\hline
 & Model & Params & FLOPs & HAM-10000 top-1-acc & Brain Tumor top-1-acc \\ 
\hline
\multirow{5}{*}{ConvNets} & MobileNetV2 \cite{sandler_mobilenetv2_2018} & 2.2M & 0.32G & 90.70\% & 99.62\% \\ 
 & MobileNetV3-L \cite{howard_searching_2019} & 4.2M & 0.23G & 90.46\% & 99.77\% \\ 
 & EfficientNetB0 \cite{tan_efficientnet_2019} & 4.0M & 0.41G & 90.70\% & 99.77\% \\ 
 & RegNetY-4.0 \cite{radosavovic_designing_2020} & 19.6M & 4.0G & 89.73\% & 99.62\% \\
 & Teacher model (ours) & 3.95M & 5.88G & 91.43\% & 99.69\% \\
 \hline
 \multirow{3}{*}{ViTs} & PVT-Tiny \cite{wang_pyramid_2021}& 12.7M & 1.88G & 87.80\% & 98.86\% \\
 & Crossvit-Ti \cite{chen_crossvit_2021}& 6.7M & 1.6G & 87.44\% & 99.08\% \\
 & CSKD-Tiny \cite{zhao_cumulative_2023} & 5.5M & 1.1G & 89.86\% & 99.24\% \\
 \hline
 \multirow{5}{*}{Hybird Models} & MobileViT-S \cite{mehta_mobilevit_2021}& 5.0M & 2.0G & 89.73\% & 99.62\% \\
 & PVTv2-B0 \cite{wang_pvt_2022} & 3.4M & 0.54G & 88.89\%  & 99.08\% \\
 & CoatNet-0 \cite{dai_coatnet_2021} & 26M & 4.21G & 90.22\% & 99.85\% \\
 & Student w/o distillation (ours) & 3.7M & 2.08G & 89.86\% & 99.77\% \\
 & HDKD (ours) & 3.7M & 2.09G & 92.15\% & 99.85\% \\
 \hline

\end{tabular}
\end{table*}

\begin{table*}[!tb]
\centering
\caption{Comparison between the teacher, student and HDKD models in terms of Precision, Recall and AUC scores on HAM-10000 dataset.}
\begin{tabular}{lcccccccccc}
\toprule
\multirow{2}{*}{Class} & \multicolumn{3}{c}{Precision} & \multicolumn{3}{c}{Recall} & \multicolumn{3}{c}{AUC} & \multirow{2}{*}{\#} \\
\cmidrule(lr){2-4} \cmidrule(lr){5-7} \cmidrule(lr){8-10}
& Teacher & Student & HDKD & Teacher & Student & HDKD & Teacher & Student & HDKD & \\
\midrule
AKIEC & 66.7\% & 54.5\% & 68.8\% & 52.2\% & 26.1\% & 47.8\% & 97.5\% & 95.6\% & 96.5\% & 23 \\
BCC & 69.0\% & 72.7\% & 63.2\% & 76.9\% & 61.5\% & 92.3\% & 98.4\% & 98.5\% & 98.6\% & 26 \\
BKL & 80.0\% & 68.3\% & 74.6\% & 65.2\% & 65.2\% & 75.8\% & 96.2\% & 93.2\% & 96.3\% & 66 \\
DF & 50.0\% & 20.0\% & 75.0\% & 33.3\% & 16.7\% & 100.0\% & 98.5\% & 95.6\% & 99.8\% & 6 \\
MEL & 66.7\% & 58.6\% & 73.1\% & 58.8\% & 50.0\% & 55.9\% & 96.4\% & 95.6\% & 96.1\% & 34 \\
NV & 95.3\% & 94.5\% & 97.4\% & 98.2\% & 98.3\% & 97.1\% & 97.7\% & 97.4\% & 98.4\% & 663 \\
VASC & 90.0\% & 100.0\% & 75.0\% & 90.0\% & 80.0\% & 90.0\% & 99.8\% & 99.7\% & 99.9\% & 10 \\
\midrule
Avg. & 73.9\% & 66.9\% & \textbf{75.3\%} & 67.8\% & 56.8\% & \textbf{79.8\%} & 97.8\% & 96.5\% & \textbf{98.0\%} & 828 \\
\midrule
W.Avg & 90.9\% & 88.7\% & \textbf{92.3\%} & 91.4\% & 89.7\% & \textbf{92.1\%} & 99.3\% & 99.0\% & \textbf{99.3\%} & 828 \\
\bottomrule
\end{tabular}
\end{table*}

\begin{table*}[!tb]
\centering
\caption{Comparison between the teacher, student, and HDKD models in terms of Precision, Recall, and AUC scores on the Brain Tumor dataset.}
\begin{tabular}{lcccccccccc}
\toprule
\multirow{2}{*}{Class} & \multicolumn{3}{c}{Precision} & \multicolumn{3}{c}{Recall} & \multicolumn{3}{c}{AUC} & \multirow{2}{*}{\#} \\
\cmidrule(lr){2-4} \cmidrule(lr){5-7} \cmidrule(lr){8-10}
& Teacher & Student & HDKD & Teacher & Student & HDKD & Teacher & Student & HDKD & \\
\midrule
Glioma      & 100\%  & 100\%  & 100\%  & 99.3\% & 99.7\% & 100\%  & 100\%  & 99.8\% & 100\%  & 300 \\
Meningioma  & 98.7\% & 99.0\% & 99.7\% & 99.7\% & 100\%  & 99.7\% & 99.9\% & 99.9\% & 99.9\% & 306 \\
No tumor    & 99.8\% & 100\%  & 99.8\% & 100\%  & 100\%  & 100\%  & 100\%  & 100\%  & 100\%  & 405 \\
Pituitary   & 100\%  & 100\%  & 100\%  & 99.3\% & 99.3\% & 99.7\% & 100\%  & 100\%  & 100\%  & 300 \\
\midrule
Avg.        & 99.6\% & 99.8\% & \textbf{99.9\%} & 99.6\% & 99.7\% & \textbf{99.8\%} & \textbf{100\%}  & 99.9\% & \textbf{100\%}  & 1311 \\
\midrule
W.Avg       & 99.6\% & \textbf{99.8\%} & \textbf{99.8\%} & 99.6\% & \textbf{99.8\%} & \textbf{99.8\%} & \textbf{100\%}  & 99.9\% & \textbf{100\%}  & 1311 \\
\bottomrule
\end{tabular}
\end{table*}

\begin{table*}[!tb]
\centering
\caption{Comparison between our HDKD model with EfficientNetB0 and CoAtNet-0 in terms of Precision, Recall, and AUC scores on the HAM-10000 dataset.}
\begin{tabular}{lcccccccccc}
\toprule
\multirow{2}{*}{Class} & \multicolumn{3}{c}{Precision} & \multicolumn{3}{c}{Recall} & \multicolumn{3}{c}{AUC} & \multirow{2}{*}{\#} \\
\cmidrule(lr){2-4} \cmidrule(lr){5-7} \cmidrule(lr){8-10}
& \cite{tan_efficientnet_2019} & \cite{dai_coatnet_2021} & HDKD & \cite{tan_efficientnet_2019} & \cite{dai_coatnet_2021} & HDKD & \cite{tan_efficientnet_2019} & \cite{dai_coatnet_2021} & HDKD & \\
\midrule
AKIEC      & 56.0\% & 100\%  & 68.8\% & 60.9\% & 39.1\% & 47.8\% & 98.1\% & 96.6\% & 96.5\% & 23 \\
BCC        & 68.0\% & 62.1\% & 63.2\% & 65.4\% & 69.2\% & 92.3\% & 98.7\% & 98.7\% & 98.6\% & 26 \\
BKL        & 68.3\% & 75.0\% & 74.6\% & 65.2\% & 68.2\% & 75.8\% & 93.0\% & 94.6\% & 96.3\% & 66 \\
DF         & 50.0\% & 75.0\% & 75.0\% & 66.7\% & 50.0\% & 100\%  & 95.7\% & 95.7\% & 99.8\% & 6  \\
MEL        & 69.0\% & 44.4\% & 73.1\% & 58.8\% & 47.1\% & 55.9\% & 96.0\% & 94.2\% & 96.1\% & 34 \\
NV         & 96.6\% & 95.0\% & 97.4\% & 97.1\% & 97.7\% & 97.1\% & 97.5\% & 96.4\% & 98.4\% & 663 \\
VASC       & 81.8\% & 100\%  & 75.0\% & 90.0\% & 80.0\% & 90.0\% & 98.4\% & 100\%  & 99.9\% & 10  \\
\midrule
Avg.       & 69.9\% & \textbf{78.8\%} & 75.3\% & 72.0\% & 64.5\% & \textbf{79.8\%} & 96.8\% & 96.6\% & \textbf{98.0\%} & 828 \\
\midrule
W.Avg      & 90.6\% & 90.4\% & \textbf{92.3\%} & 90.7\% & 90.2\% & \textbf{92.1\%} & 99.1\% & 99.0\% & \textbf{99.3\%} & 828 \\
\bottomrule
\end{tabular}
\end{table*}

\begin{table*}[!tb]
\centering
\caption{Comparison between our HDKD model with EfficientNetB0 and CoAtNet-0 in terms of Precision, Recall, and AUC scores on the Brain Tumor dataset.}
\begin{tabular}{lcccccccccc}
\toprule
\multirow{2}{*}{Class} & \multicolumn{3}{c}{Precision} & \multicolumn{3}{c}{Recall} & \multicolumn{3}{c}{AUC} & \multirow{2}{*}{\#} \\
\cmidrule(lr){2-4} \cmidrule(lr){5-7} \cmidrule(lr){8-10}
& \cite{tan_efficientnet_2019} & \cite{dai_coatnet_2021} & HDKD & \cite{tan_efficientnet_2019} & \cite{dai_coatnet_2021} & HDKD & \cite{tan_efficientnet_2019} & \cite{dai_coatnet_2021} & HDKD & \\
\midrule
Glioma      & 100\% & 100\% & 100\% & 99.7\% & 99.7\% & 100\% & 100\% & 100\% & 100\% & 300 \\
Meningioma  & 99.3\% & 99.4\% & 99.7\% & 99.7\% & 100\% & 99.7\% & 99.9\% & 100\% & 99.9\% & 306 \\
No tumor    & 99.8\% & 100\% & 99.8\% & 100\% & 100\% & 100\% & 100\% & 100\% & 100\% & 405 \\
Pituitary   & 100\% & 100\% & 100\% & 99.7\% & 99.7\% & 99.7\% & 100\% & 100\% & 100\% & 300 \\
\midrule
Avg.        & 99.8\% & 99.8\% & \textbf{99.9\%} & \textbf{99.8\%} & \textbf{99.8\%} & \textbf{99.8\%} & \textbf{100\%} & \textbf{100\%} & \textbf{100\%} & 1311 \\
\midrule
W.Avg       & \textbf{99.8\%} & \textbf{99.8\%} & \textbf{99.8\%} & \textbf{99.8\%} & \textbf{99.8\%} & \textbf{99.8\%} & \textbf{100\%} & \textbf{100\%} & \textbf{100\%} & 1311 \\
\bottomrule
\end{tabular}
\end{table*}

\section{Ablation Studies}\label{sec5}
In this section, we compare the performance of our MBCSA block against MBConv block. In addition, we explore the effect of increasing the stride from 2 to 4 in the initial stem stage, observing a trade-off between performance and efficiency. Lastly, we conduct feature visualization analysis to observe the effect of distillation on the student model. \\
\newline
\textbf{Comparison between MBCSA and MBConv blocks.} The MBConv block is compared with MBCSA block when incorporating them as the main convolutional block in the teacher, student without distillation and HDKD models. As observed from Table 10, MBCSA block improves the teacher model performance across both datasets, this is attributed to its utilization of spatial and channel attention from the CBAM module. However, in the hybrid student model, which integrates transformer layers that are mainly reliant on attention, the enhancement of MBCSA may diminish. 
\newline
 \indent As observed in Table 10, the teacher model showed a significant improvement with the MBCSA block, especially on the HAM-10000 dataset. This improvement in the teacher model translates to better knowledge transfer to the student model using distillation (HDKD). As a result, the HDKD model achieved about 1\% higher accuracy on the HAM-10000 dataset when using the MBCSA block compared to the MBConv block. Furthermore, both blocks exhibit similar complexity in terms of parameters and FLOPs.
\newline
 \indent Additionally, to further demonstrate the efficiency of MBCSA block, we replaced the mobile convolution blocks in state-of-the-art lightweight models such as MobileNetV3 \cite{howard_searching_2019}, MobileViT \cite{mehta_mobilevit_2021} and Coatnet \cite{dai_coatnet_2021} with MBCSA block and evaluated them on HAM-10000 dataset in both setups. The inclusion of the MBCSA block led to accuracy improvements on the HAM-10000 dataset of 0.84\% and 0.85\% in MobileNetV3 and MobileViT, respectively as given in Table 11. This reveals that the incorporation of MBCSA blocks in different state-of-the-art models improves the accuracy while having approximately the same computational requirements.
\begin{table*}[!tb]
\centering
\caption{Comparison between MBSCA and MBConv blocks. We compare our teacher model, student model (i.e., without distillation) and HDKD models in terms of performance and complexity, when either incorporating MBCSA or MBConv block.}
\begin{tabular}{>{\centering\arraybackslash}m{3cm} c c c c c} 
 \toprule
\vspace{0.8\baselineskip} 
  \makecell{\vspace{0.2cm} Model} & Block & Params & FLOPs & \makecell{HAM-10000 \\ Dataset Accuracy} & \makecell{Brain Tumor \\ Dataset Accuracy}\\ 
 \toprule
\multirow{2}{*}{\makecell{Teacher \\ model}} & MBConv & 3.95M & 5.85G & 90.58\% & 99.62\% \\ 
 & MBCSA & 3.95M & 5.88G & 91.43\% & 99.69\% \\ 
 \midrule
 \multirow{2}{*}{\makecell{Student (w/o) \\ distill model}} & MBConv & 3.72M & 2.07G & 89.86\% & 99.77\% \\ 
 & MBCSA & 3.72M & 2.08G & 89.86\% & 99.77\% \\ 
 \midrule
 \multirow{2}{*}{\makecell{HDKD \\ model}} & MBConv & 3.73M & 2.08G & 91.18\% & 99.85\% \\ 
 & MBCSA & 3.73M & 2.09G & 92.15\% & 99.85\% \\ 
 \bottomrule
\end{tabular}
\end{table*}
\begin{table*}[!tb]
\centering
\caption{Comparison between MobileNetV3, MobileViT and Coatnet models when utilizing their main mobile convolution blocks and when utilizing MBCSA block instead.}
\begin{tabular}{lcccc}
\toprule
Model       & Block   & Params  & FLOPs  & HAM-10000 Dataset Accuracy \\
\midrule
\multirow{2}{*}{\makecell{MobileNetV3 \\ model}} & MBConv  & 4.2M    & 0.23G  & 90.46\%         \\
       & MBCSA   & 4.2M    & 0.23G  & 91.30\%         \\
\midrule
\multirow{2}{*}{\makecell{MobileViT \\  model}}  & MBConv  & 5.0M    & 2.0G   & 89.73\%         \\
       & MBCSA   & 5.1M    & 2.0G   & 90.58\%         \\
\midrule
\multirow{2}{*}{\makecell{CoAtNet \\  model}}    & MBConv  & 26M     & 4.21G  & 90.22\%         \\
       & MBCSA   & 26M     & 4.22G  & 90.22\%         \\
\bottomrule
\end{tabular}
\end{table*}

\\
\newline
\noindent \textbf{Effect of stride in the initial stem stage.} Utilizing a larger stride in the initial stem stage significantly reduces FLOPs and latency, due to operating on smaller feature maps in all subsequent stages. We aim to investigate the effect of increasing the stride from 2 to 8 in the stem stage to observe changes in model performance and complexity. As observed in Table 12, the accuracy of both the teacher and student models  decreases when the stride value increases, particularly on the HAM-10000 dataset. Conversely, FLOPs is reduced by approximately 4\(\times\) and 16\(\times\) for both models when utilizing stride of 4 and 8, respectively, due to processing lower-resolution feature maps across all stages, which decreases the overall number of operations required. Thus, choosing a larger stride is beneficial for enhancing inference time, making it advantageous for deployment on edge devices.
\begin{table*}[!tb]
\centering
\caption{Comparison of different stride values in stem stage. We compare our teacher model, student model (i.e., without distillation) and HDKD when using different stride values (i.e., 2, 4 and 8).}
\begin{tabular}{>{\centering\arraybackslash}m{3cm} c c c c c c} 
\toprule
\vspace{0.8\baselineskip} 
\makecell{\vspace{0.2cm} Model} & \makecell{Stride in \\ stem stage} & Params & FLOPs & Latency (ms) & \makecell{HAM-10000 \\ Dataset Accuracy} & \makecell{Brain Tumor \\ Dataset Accuracy} \\ 
\toprule
\multirow{3}{*}{\makecell{Teacher \\ model}} 
& 2 & 3.95M & 5.88G & 112.9 & 91.43\% & 99.69\% \\ 
& 4 & 3.95M & 1.47G & 51.5  & 90.58\% & 99.69\% \\ 
& 8 & 3.95M & 0.368G & 25.8  & 89.61\% & 99.16\% \\ 
\midrule
\multirow{3}{*}{\makecell{Student (w/o) \\ distill model}} 
& 2 & 3.72M & 2.08G & 38.7  & 89.86\% & 99.77\% \\ 
& 4 & 3.69M & 0.511G & 22.0  & 89.25\% & 99.47\% \\ 
& 8 & 3.68M & 0.139G & 12.8  & 88.29\% & 98.78\% \\ 
\midrule
\multirow{3}{*}{\makecell{HDKD \\ model}} 
& 2 & 3.73M & 2.09G & 39.8  & 92.15\% & 99.85\% \\ 
& 4 & 3.69M & 0.513G & 22.5  & 91.06\% & 99.77\% \\ 
& 8 & 3.68M & 0.141G & 13.9  & 89.73\% & 99.31\% \\ 
\bottomrule
\end{tabular}
\end{table*}
 
\\
\newline
\noindent \textbf{Feature visualization analysis.} To explore how distillation impacts HDKD activation maps, we provide visual comparisons of the activation maps generated at the third stage for our three models: the teacher model, the student model without distillation, and the HDKD model. The visual comparison shown in Figure 5. reveals that the HDKD model's activation maps effectively highlight infection areas more than those of the teacher model and the student model without distillation. Moreover, due to feature distillation, the activation maps of the HDKD model closely resemble those of the teacher model, but with enhanced detail. This demonstrates the HDKD model's superior ability to capture salient features and its ability to generalize across various tasks.

\begin{figure}[H]
\centering
    \includegraphics[width=0.44\linewidth]{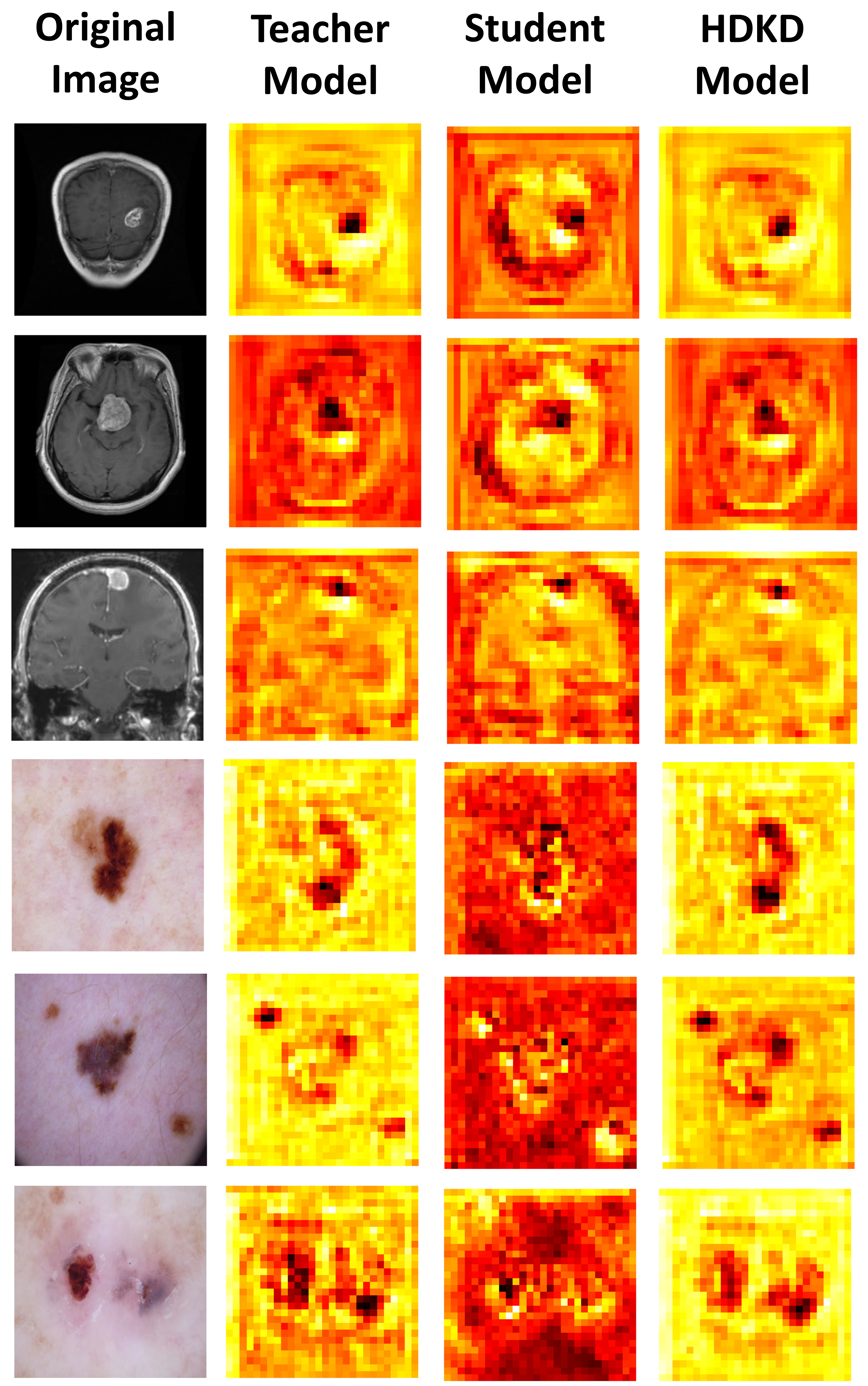}
    \caption{\textbf{Feature visualization analysis.} We visualize the activation maps of the last block in the final convolutional stage (3\textsuperscript{rd} stage) for each of the teacher model, student model without distillation, and HDKD model. These activation maps are obtained by averaging across the channel dimension, resulting in a single global channel of size 28\(\times\)28.}
\end{figure}

\section*{Conclusions}\label{sec6}
Knowledge distillation has emerged as a powerful technique in various deep learning applications, offering enhancements in model performance, size compression, and effective generalization with limited data. However, Knowledge distillation faces limitations, particularly concerning misalignment issues between teacher and student models when employing feature-based knowledge distillation techniques. This is especially prevalent in scenarios where Convolutional Neural Networks (CNNs) are distilled to Vision Transformers (ViTs) due to the structural difference between these models which can lead to sub-optimal knowledge transfer and additional operations overhead. In this paper, we introduce an efficient teacher-student paradigm called Hybrid Data-Efficient Knowledge Distillation (HDKD). HDKD enables the efficient knowledge transfer from the teacher model to the student model without any additional alignment operations. We introduce an efficient MBCSA block as the main convolutional block in the proposed teacher and student models. The teacher model is designed as a CNN network while the student model is designed as a hybrid network to combine the benefits of both CNNs (inductive biases) and ViTs (global processing). The teacher and student models share the same convolutional structure which allows applying direct distillation between them without any alignment operations or information loss. Extensive experiments on Brain Tumor and HAM-10000 datasets demonstrate the superiority and generalization of HDKD compared to the state-of-the-art models while also maintaining a light-weight design, enabling its deployment on resource-constrained devices. While the study focused on applying HDKD on different medical tasks, we plan in the future to experiment our proposed models on other domains to demonstrate its applicability to different AI and engineering areas.

\bibliographystyle{ieeetr} 
\bibliography{references} 

\end{document}